# Strong-Coupling Superconductivity with $T_c$ ~ 10.8 K Induced by P doping in the Topological Semimetal Mo$_5$Si$_3$

Bin-Bin Ruan[1*†], Jun-Nan Sun[1,2,3†], Yin Chen[1,4], Qing-Song Yang[1,5], Kang Zhao[1,6], Meng-Hu Zhou[1], Ya-Dong Gu[1,5], Ming-Wei Ma[1,7], Gen-Fu Chen[1,5,7], Lei Shan[2,3*] and Zhi-An Ren[1,5*]

[1] Institute of Physics and Beijing National Laboratory for Condensed Matter Physics, Chinese Academy of Sciences, Beijing 100190, China
[2] Information Materials and Intelligent Sensing Laboratory of Anhui Province, Institutes of Physical Science and Information Technology, Anhui University, Hefei 230601, China
[3] Key Laboratory of Structure and Functional Regulation of Hybrid Materials (Anhui University), Ministry of Education, Hefei 230601, China
[4] School of Materials Science and Engineering, University of Science and Technology Beijing, Beijing 100083, China
[5] School of Physical Sciences, University of Chinese Academy of Sciences, Beijing 100049, China
[6] Pilot National Laboratory for Marine Science and Technology, Qingdao, Shandong 266237, China
[7] Songshan Lake Materials Laboratory, Dongguan, Guangdong 523808, China

**ABSTRACT**

By performing P doping on the Si sites in the topological semimetal Mo$_5$Si$_3$, we discover strong-coupling superconductivity in Mo$_5$Si$_{3-x}$P$_x$ ($0.5 \leq x \leq 2.0$). Mo$_5$Si$_3$ crystallizes in the W$_5$Si$_3$-type structure with space group of $I4/mcm$ (No. 140), and is not a superconductor itself. Upon P doping, the lattice parameter $a$ decreases while $c$ increases monotonously. Bulk superconductivity is revealed in Mo$_5$Si$_{3-x}$P$_x$ ($0.5 \leq x \leq 2.0$) from resistivity, magnetization, and heat capacity measurements. $T_c$ in Mo$_5$Si$_{1.5}$P$_{1.5}$ reaches as high as 10.8 K, setting a new record among the W$_5$Si$_3$-type superconductors. The upper and lower critical fields for Mo$_5$Si$_{1.5}$P$_{1.5}$ are 14.56 T and 105 mT, respectively. Moreover, Mo$_5$Si$_{1.5}$P$_{1.5}$ is found to be a fully gapped superconductor with strong electron-phonon coupling. First-principles calculations suggest that the enhancement of electron-phonon coupling is possibly due to the shift of the Fermi level, which is induced by electron doping. The calculations also reveal the nontrivial band topology in Mo$_5$Si$_3$. The $T_c$ and upper critical field in Mo$_5$Si$_{3-x}$P$_x$ are fairly high among pseudobinary compounds. Both of them are higher than those in NbTi, making future applications promising. Our results suggest that the W$_5$Si$_3$-type compounds are ideal platforms to search for new superconductors. By examinations of their band topologies, more candidates for topological superconductors can be expected in this structural family.

---

\* Corresponding authors (email: bbruan@mail.ustc.edu.cn; lshan@ahu.edu.cn; renzhian@iphy.ac.cn)
† These authors contributed equally to this work.

## INTRODUCTION

Topological superconductors, hosting both gapped bulk superconducting states and gapless surface states, have attracted much attention in recent years. The most fascinating feature in a topological superconductor is that its quasiparticle excitations form Majorana zero modes (MZMs) at boundaries and vortices [1-3]. The MZMs obey the non-Abelian statistics, and are thus suitable for the realization of fault-tolerant quantum computations [4, 5].

In view of the above intriguing merits, much effort has been devoted to seek for topological superconductivity in real materials. One approach is to search for superconductors with odd parity, as demonstrated in $Sr_2RuO_4$, $Sn_{1-x}In_xTe$, or $T_d$-$MoTe_2$ [6-9]. However, odd-parity superconductors are very rare, and their superconductivity is generally fragile to impurities or disorders. Moreover, all these superconductors have very low (< 2 K) superconducting transition temperatures ($T_c$s), limiting possible applications. Another approach, as proposed by Fu and Kane [10], is to fabricate heterostructures made of superconductors and topological insulators, where the topological surface states (TSSs) become superconducting from the proximity effect. The realizations [11-13], however, require delicate device fabrications and face the challenges of lattice mismatch and interface complexity.

Consequently, researchers in this field have been moving much of their attention to find bulk superconductors that also possess nontrivial band topologies. Such conception is simple but effective. In this spirit, superconducting topological materials such as $\beta$-$PdBi_2$, doped $Bi_2Se_3$, PdBi, 2M-$WS_2$ were proposed to be candidates for topological superconductors [14-21]. In many of them, spectroscopy methods such as scanning tunneling microscopy (STM) and angle-resolved photoemission spectroscopy (ARPES) successfully confirmed the existence of TSSs [20-23], which in turn verified the effectiveness of the conception.

Recently, superconductivity was observed in Re doped $Mo_5Si_3$, with a maximal $T_c$ of 5.8 K in $Mo_3Re_2Si_3$ [24]. Not only did Ref. 24 set a record-high $T_c$ in $W_5Si_3$-type superconductors, it also emphasized the nontrivial band topology, making $Mo_3Re_2Si_3$ a candidate for topological superconductors. We noticed that, just like the cases of Cu/Sr/Nb doped $Bi_2Se_3$ [14-17, 19], superconductivity was successfully induced by carrier doping in topological material $Mo_5Si_3$. We thus systematically examined the doping effects not only on the Mo sites, but also on the Si sites in $Mo_5Si_3$.

In this paper, we report detailed characterizations of $Mo_5Si_{3-x}P_x$ ($0 \leq x \leq 2.0$), in which bulk superconductivity with $T_c$ as high as 10.8 K is observed. In addition, $Mo_5Si_2P$ and $Mo_5Si_{1.5}P_{1.5}$ are found to host strong electron-phonon coupling. The enhancement of the coupling strength is due to the shift of the Fermi level, and possibly the phonon softening, as revealed by the heat capacity measurements and first-principles calculations. A series of superconducting parameters for $Mo_5Si_{1.5}P_{1.5}$ are determined, and the electronic band topologies are briefly discussed.

## METHODS

Polycrystalline samples of $Mo_5Si_{3-x}P_x$ ($x = 0, 0.5, 1.0, 1.2, 1.3, 1.5, 1.6,$ and $2.0$) were prepared by solid state reaction. Elements of Mo (99.9%, powder), Si (99.999%, powder), and P (99.99%, powder) were mixed thoroughly before pressed into pellets. The pellets were placed into alumina crucibles before sealed into silica tubes under argon. The tubes were slowly heated to 1073 K and held for 24 hours. Then the products were thoroughly ground, pressed into pellets, put into alumina crucibles, and sealed into tantalum tubes under argon. The tubes were heated under high-purity argon at 1923 K for 20 hours. All the manipulations except sealing and heating were carried out in a glove box filled with high-purity argon. The final products showed silver metallic lusters and were stable in air.

The room temperature powder x-ray diffraction (XRD) data were collected on a PAN-analytical x-ray diffractometer with Cu-$K_\alpha$ radiation. Rietveld refinements were carried out using the GSAS package [25]. The resistivity and heat capacity data were collected on a physical property measurement system (PPMS, Quantum Design). The magnetization measurements were performed on a magnetic property measurement system (MPMS, Quantum Design). The chemical compositions were determined by an energy-dispersive x-ray (EDX) spectrometer equipped on a Phenom ProX scanning electron microscope. More details about the measurements can be found in our previous study [26].

First-principles calculations were performed based on the density functional theory (DFT), as implemented in the QUANTUM ESPRESSO package [27]. The exchange-correlation functionals of PBE based on the generalized gradient approximation (GGA) were chosen. The optimized normconserving pseudopotentials [28] were used. Before the calculations for charge densities, the lattice parameters, as well as the atomic positions, were fully relaxed until the force on each atom was less than 0.0001 Ry/Bohr. A Monkhorst–Pack grid of 15×15×11 was applied in the self-consistent calculations. P doping on the Si sites was treated by the method of virtual crystal approximation (VCA). The validation of VCA has been checked with the supercell results. (see Supplementary Information)

## RESULTS
### Structural characterizations

Figure 1(a) demonstrates the crystal structure of $Mo_5Si_{3-x}P_x$, one may notice that the structure features Si–Si chains along the $c$ axis. XRD patterns of polycrystalline $Mo_5Si_{3-x}P_x$ ($0 \leq x \leq 2.0$) are shown in Figure 1(e). Without P doping, a phase pure $Mo_5Si_3$ sample, which is of the tetragonal $W_5Si_3$ type (space group $I4/mcm$), is successfully obtained. Upon doping, an impurity phase of $Mo_3P$ emerges. Using MoP precursor instead of P in the preparation procedure was found to be beneficial to reduce the amount of $Mo_3P$ in the final products. However, we were not able to completely remove the $Mo_3P$ impurity. This is possibly due to the inevitable evaporation of P at high temperature. As a result, the actual contents of P in the products should be less than the nominal ones, which was confirmed by our EDX measurements. The measured values of $x$ are listed in Table 1, and are plotted in Figure 1(d). Notice that the measured P contents are not far from the nominal ones. For simplification, $x$ in $Mo_5Si_{3-x}P_x$ mentioned hereafter represent the nominal values.

The diffraction peaks of $Mo_5Si_{3-x}P_x$ evidently shift with $x$ increasing, as shown in Figure 1(f). To gain insights into the crystallographic parameters, we performed Rietveld refinements to each XRD pattern. Details of the refinement results are listed in Table 1. Two of them ($x = 0.5$ and $x = 1.5$) are shown as examples in Figure 1(b). The small values of $R_p$, $R_{wp}$, and $\chi^2$ suggest the refinements are satisfactory. As shown in Figure 1(c), the $a$-axis shrinks while the $c$-axis expands monotonously, indicating a successful P doping into $Mo_5Si_3$. This doping behavior is different from that in $Mo_{5-x}Re_xSi_3$, where only the $a$-axis was changed [24]. We note that the shrinkage of $a$-axis of $Mo_5SiP_2$ compared with $Mo_5Si_3$ is around 1.6%, while the changes of the lattice parameters of $Mo_3P$ impurity are less than 0.1%. This means that the Si doping content in $Mo_3P$ is insignificant (if not zero) in our samples. There are two different Wyckoff positions (Si1 at $4a$ and Si2 at $8h$) of Si in $Mo_5Si_3$. Therefore, there could be a site-selection in P doping. We carefully examined the evolution of all the bond lengths in $Mo_5Si_{3-x}P_x$. However, no evidence backing this assumption was found. No reflections from supercell were observed in the XRD patterns either. P is thus believed to randomly take all the Si sites. (We should note that it is generally very difficult to distinguish P from Si by XRD measurements, so chances are that there still exists nonequivalent doping between the Si1 and Si2 sites.) The shrinkage of $a$ and expansion of $c$ are consistent with our DFT relaxation results. (see Supplementary Information)

### Superconducting properties

The temperature dependence of electrical resistivity ($\rho$) for $Mo_5Si_{3-x}P_x$ ($0 \leq x \leq 2.0$) is shown in Figure 2(a). All of the samples show metallic behaviors. For the undoped sample $Mo_5Si_3$, $\rho$ reads ~ 0.22 mΩ cm at 300 K and decreases monotonously with the decrease of temperature. $\rho(T)$ for $Mo_5Si_3$ shows no superconducting transitions down to 1.8 K. These results are in good agreement with those in the references [24, 29], where no superconductivity was observed above 0.15 K. P doping into $Mo_5Si_3$ introduces superconductivity, as revealed by abrupt drops of $\rho(T)$ curves for the doped samples. The region of the superconducting transitions is emphasized in Figure 2(b). For the samples with $x \geq 1.0$, the normal state $\rho(T)$ curves obviously show upwards concave features, similar to those observed in the $A$15 compounds, which can be interpreted by a parallel-resistor model [30].

To investigate the magnetic properties of the superconducting samples, the DC magnetic susceptibility ($4\pi\chi$) of $Mo_5Si_{3-x}P_x$ ($0.5 \leq x \leq 2.0$) was measured and is shown in Figure 2(c). Notice the data have been corrected with the corresponding demagnetization factors. In the zero-field-cooling (ZFC) run, $4\pi\chi$ of each sample quickly approaches a constant at low temperature, indicating the occurrence of superconductivity. The shielding fractions are close to or larger than 100%, validating bulk superconductivity in $Mo_5Si_{3-x}P_x$. $T_c$s can be determined from the onset temperature to deviate from the normal states, which are in good agreements with those obtained from the $\rho(T)$ data. It should be mentioned that the diamagnetic signals of $Mo_3P$ ($T_c = 5.6$ K [31]) in the $4\pi\chi(T)$ curves are negligible (it is obvious only in the heavily doped sample $Mo_5SiP_2$). This is presumably due to the shielding effects of $Mo_5Si_{3-x}P_x$, which have higher $T_c$s than $Mo_3P$. The absolute values of $4\pi\chi$ in the field-cooling (FC) runs are significantly lower than those in the ZFC runs, indicating large pinning effects in the superconducting samples.

Gathering the data from $\rho(T)$ and $4\pi\chi(T)$, we are able to conclude the evolution of $T_c$ in $Mo_5Si_{3-x}P_x$ ($0 \leq x \leq 2.0$), as listed in Table 1, and shown in Figure 2(d). Non-superconducting $Mo_5Si_3$ becomes superconducting with P doping. $T_c$ quickly increases, exceeding 10 K in $Mo_5Si_{1.8}P_{1.2}$, while further P doping brings up little change in $T_c$. Simultaneously, the residual resistivity ratio (RRR) decreases regularly upon P doping, which is reasonable since P doping introduces more defects in the sample.

Now we move on to the discussion of the superconducting nature by conducting a detailed investigation on the $Mo_5Si_{1.5}P_{1.5}$ sample. We choose to characterize this sample in detail because it hosts the highest $T_c$, and has less $Mo_3P$ impurities compared with heavier-doped samples. $\rho(T)$ of $Mo_5Si_{1.5}P_{1.5}$ under various magnetic fields ($\mu_0H = 0 - 15.5$ T) is shown in Figure 3(a). Under zero magnetic field, $\rho(T)$ starts to drop abruptly at a $T_c^{onset}$ of 10.80 K, and reaches zero at a $T_c^{zero}$ of 10.70 K, resulting in a superconducting transition width of only 0.1 K. Upon the application of magnetic field, superconductivity in $Mo_5Si_{1.5}P_{1.5}$ is gradually suppressed. $T_c$ under different magnetic fields is determined by the 50% criterion, $i.e.$ the temperature where $\rho(T)$ reaches 50% of that of the normal state. The phase diagram of upper critical field ($\mu_0H_{c2}$) versus $T$ is therefore plotted in Figure 3(d). It can be seen that the Ginzburg–Landau (G–L) model $\mu_0H_{c2}(T) = \mu_0H_{c2}(0)[1 - (T/T_c)^2]/[1 + (T/T_c)^2]$ gives a satisfying fit of the experimental results in the whole temperature range. $\mu_0H_{c2}(0)$ is fitted to be 14.56 T, which is lower than the Pauli paramagnetic limit (20.0 T).

We have also performed isothermal magnetization measurements on $Mo_5Si_{1.5}P_{1.5}$. The hysteresis loop at 2 K is shown in Figure 3(c). A typical behavior of a type-II superconductor is observed. The isothermal magnetization curves under various temperatures are shown in Figure 3(b), from which the lower critical field ($\mu_0H_{c1}$) can be determined from the deviation of the curves from the initial Meissner states. $\mu_0H_{c1}$ under different temperature is plotted in Figure 3(d). One can easily fit $\mu_0H_{c1}(T)$ with the well-known G–L expression: $\mu_0H_{c1}(T) = \mu_0H_{c1}(0)(1 - t^2)$, where $t$ is the normalized temperature $T/T_c$. The fit gives a $\mu_0H_{c1}(0)$ of 105 mT.

Based on the results of $\mu_0H_{c1}(0)$ and $\mu_0H_{c2}(0)$, a series of superconducting parameters can be obtained. The G–L coherence length ($\xi_{GL}$) of $Mo_5Si_{1.5}P_{1.5}$ is calculated to be 4.75 nm by the relation: $\mu_0H_{c2}(0) = \Phi_0/(2\pi\xi_{GL}^2)$, in which $\Phi_0$ stands for the magnetic flux quantum. The superconducting penetration depth ($\lambda_{GL}$) is calculated by $\mu_0H_{c1}(0) = \Phi_0/(4\pi\lambda_{GL}^2)(\ln\kappa + 0.5)$, where $\kappa \equiv \lambda_{GL}/\xi_{GL}$ is the G–L parameter. Consequently, we obtain $\lambda_{GL} = 70.5$ nm and $\kappa = 14.8$. The value of $\kappa$ is far larger than $1/\sqrt{2}$, again suggesting $Mo_5Si_{1.5}P_{1.5}$ to be a type-II superconductor. The thermodynamic critical field can therefore be determined by $\mu_0H_c(0) = \mu_0\sqrt{H_{c1}(0)H_{c2}(0)/\ln\kappa}$ to be 0.75 T. All these superconducting parameters are summarized in Table 2.

In order to take more insight into the superconductivity, as well as the thermodynamic properties of $Mo_5Si_{3-x}P_x$, we measured the specific heat ($C_p$) of $Mo_5Si_{3-x}P_x$ ($x = 0, 1.0, 1.5$). As the doped samples contained superconducting

Mo$_3$P, we synthesized a phase pure Mo$_3$P sample, whose $C_p(T)$ was measured before subtracted from the raw data of Mo$_5$Si$_{3-x}$P$_x$. (see Supplementary Information) This approach is similar to that in Mo$_5$PB$_2$ [32]. The corrected data are shown in Figure 4. No superconducting transitions are observed in Mo$_5$Si$_3$, while clear anomalies are found at 8.46 K and 10.62 K for $x = 1.0$ and $x = 1.5$, respectively, evidencing the bulk superconductivity. The values of $T_c$ from $C_p$ measurements correspond well with those from the resistivity and the magnetization measurements. For each sample, the behavior of $C_p$ at the normal state up to 18 K can be well described with the Debye model: $C_p(T) = \gamma T + \beta T^3 + \delta T^5$, in which the three terms stand for the Sommerfeld term, the contributions from harmonic phonons and anharmonic phonons, respectively. The fitted curves are shown in Figure 4(a) as the dash lines. $\gamma$ and $\beta$ for each sample are listed in Table 2. Notice that from $x = 0$ to $x = 1.5$, $\gamma$ almost doubles, while $\beta$ becomes an order of magnitude larger. We calculate the Debye temperature ($\Theta_D$) by $\Theta_D = (12\pi^4 NR/5\beta)^{1/3}$, in which $N$ is the number of atoms per formula unit (f.u.), and $R$ is the ideal gas constant. The results are also listed in Table 2. One may see that P doping greatly reduces the value of $\Theta_D$ (from 619 K to 314 K), implying substantial softening of the lattice. This is quite surprising since the atomic mass of P is close to that of Si. The softening of phonons in P doped samples may be related to the emergence of phonon soft modes. Detailed theoretical calculations will help to elucidate this topic in the future.

The electronic contribution to $C_p$ can thus be obtained by subtracting the phonon terms. Temperature dependence of electronic specific heat ($C_e$) is shown in Figure 4(b). Notice that the normalized $C_e$ jumps at $T_c$ ($\Delta C_e/\gamma T_c$) are 2.03 and 2.06 for Mo$_5$Si$_2$P and Mo$_5$Si$_{1.5}$P$_{1.5}$, respectively. These values are much larger than the BCS weak coupling ratio (1.43), suggesting strong coupling in these samples.

For a strong-coupling superconductor, the electron-phonon coupling parameter ($\lambda_{ep}$) can be estimated by the McMillan formula modified by Allen and Dynes [33, 34]:

$$T_c = \frac{\omega_{ln}}{1.2} \exp\left[\frac{1.04(1+\lambda_{ep})}{\mu^*(1+0.62\lambda_{ep}) - \lambda_{ep}}\right], \quad (1)$$

where the logarithmic average phonon frequency $\omega_{ln}$ is given by [35]:

$$\frac{\Delta C_e}{\gamma T_c} = 1.43\left[1 + 53\left(\frac{T_c}{\omega_{ln}}\right)^2 \ln\left(\frac{\omega_{ln}}{3T_c}\right)\right]. \quad (2)$$

By setting the Coulomb screening parameter $\mu^* = 0.13$, we get $\lambda_{ep} = 1.12$ and 1.15 for $x = 1.0$ and $x = 1.5$, respectively. We further calculate the density of states at the Fermi level using $N(E_F) = 3\gamma/[\pi^2 k_B^2(1+\lambda_{ep})]$, where $k_B$ is the Boltzmann constant. The calculations give $N(E_F) = 4.99$ and 7.30 eV$^{-1}$ f.u.$^{-1}$ for $x = 1.0$ and $x = 1.5$, respectively.

$C_e$ in the superconducting state is treated by calculating the entropy with [36]:

$$S(T) = -\frac{3\gamma}{\pi^3 k_B}\int_0^{2\pi}\int_0^{\infty}\left[f\ln f + (1-f)\ln(1-f)\right]d\varepsilon d\varphi, \quad (3)$$

in which $f = 1/\left[1 + \exp\left(\sqrt{\varepsilon^2 + \Delta^2(\varphi, T)}/k_B T\right)\right]$ stands for the Fermi distribution of the quasiparticles. Here, we find that a conventional $s$-wave gap function reproduces the data well, and the so-called $\alpha$ model is applied. In this model, the angular independent gap function $\Delta(T)$ is expressed as $\Delta(T) = \alpha/\alpha_{BCS}\Delta_{BCS}(T)$ where $\alpha_{BCS}$ is the weak-coupling gap ratio (1.76) [37]. $C_e$ is calculated from $C_e = T\frac{\partial S}{\partial T}$. Fittings to the $C_e$ data are illustrated in Figure 4(b), from which we obtain the superconducting gap values at zero temperature $\Delta_0 = 1.48$ and 1.96 meV for $x = 1.0$ and $x = 1.5$, respectively. The coupling strength $\Delta_0/k_BT_c$ are thus estimated to be 2.03 and 2.14. Again, these values apparently exceed $\alpha_{BCS}$, evidencing strong-coupled superconductivity.

**First-principles calculations**

The results of first-principles calculations for $Mo_5Si_{3-x}P_x$ ($x$ = 0, 1.0, 1.5) are summarized in Figure 5, with the electronic band structures of $x$ = 0 (with and without spin-orbit coupling), $x$ = 1.0, and $x$ = 1.5 shown in Figure 5(a)–(d), respectively. For all these samples, there are multiple bands crossing the Fermi level ($E_F$), consistent with the metallic nature of $Mo_5Si_{3-x}P_x$. By comparing Figure 5(a) and 5(b), we conclude that the spin-orbit coupling (SOC) has negligible effects on the band structures, although it opens finite gaps on several $k$ points. The shapes of the electronic bands for $Mo_5Si_2P$ and $Mo_5Si_{1.5}P_{1.5}$ are basically the same with that of $Mo_5Si_3$, which means that the bands can be considered rigid in our case. One major feature of the band structure in Figure 5(a) is the flat band dispersions at around 0.25 eV above $E_F$ (as indicated by the shadowy box), which gradually shift to $E_F$ when $x$ increases. This process is clearly observed when we examine the evolution of the density of states (DOS) upon P doping (shown in Figure 5(g)). $E_F$ of $Mo_5Si_3$ locates in a dip of DOS, and it shifts to a major peak for $Mo_5Si_{1.5}P_{1.5}$. Theoretical values of DOS at $E_F$ ($N'(E_F)$) are 4.13, 7.24, and 8.04 eV$^{-1}$ f.u.$^{-1}$ for $x$ = 0, 1.0, and $x$ = 1.5, respectively. The trend of $N'(E_F)$ is consistent with the change of the Sommerfeld parameter $\gamma$. We notice that $N'(E_F)$ of $Mo_5Si_{1.5}P_{1.5}$ corresponds fairly well with the experimental one, while $N'(E_F)$ of $Mo_5Si_2P$ is much larger compared with $N(E_F)$. This means that $\lambda_{ep}$ of $Mo_5Si_2P$ is probably overestimated, and equation (3) is not applicable. In fact, if we use the inverted McMillan formula [34]:

$$\lambda_{ep} = \frac{1.04 + \mu^* \ln(\Theta_D/1.45T_c)}{(1-0.62\mu^*)\ln(\Theta_D/1.45T_c) - 1.04}, \quad (4)$$

$\lambda_{ep}$ and $N(E_F)$ for $Mo_5Si_2P$ are estimated to be 0.69, and 6.30 eV$^{-1}$ f.u.$^{-1}$, respectively.

We further examine the band topology of $Mo_5Si_3$. Since $Mo_5Si_3$ possesses both time-reversal symmetry and inversion symmetry, the $Z_2$ topological invariants can be easily calculated by checking the wavefunction parities on the eight time-reversal invariant $k$ points [38]. As illustrated in Figure 5(b), we calculate the $Z_2$ indexes for several bands near $E_F$. Notice that not all these $Z_2$ indexes are well-defined, since there might be no gap between one band and another. Nevertheless, SOC opens finite gaps between bands 160 and 162, and between bands 164 and 166. $Z_2$ indexes for these two gaps are (0;111) and (0;000), respectively. Our results are consistent with previous studies [24, 39], suggesting that bulk $Mo_5Si_3$ (with SOC) falls into the weak topological insulator state. Unfortunately, this means that the TSSs in $Mo_5Si_3$ are fragile, and are unlikely to survive with P doping, which inevitably introduces defects. What makes it worse is that $E_F$ shifts to higher energies in the doped samples, which could push the system into a topologically trivial state. In a word, P doping into $Mo_5Si_3$ may destroy its TSSs, making it less possible to realize topological superconductivity in $Mo_5Si_{3-x}P_x$.

**DISCUSSION**

In McMillan's formalism [34], the electron-phonon coupling strength is given by $\lambda_{ep} = [N(E_F)\langle I^2\rangle]/[M\langle\omega^2\rangle]$, where $M$ stands for the atomic mass, $\langle I^2\rangle$ and $\langle\omega^2\rangle$ are averages of the squared electronic matrix elements on the Fermi surface, and of the squared phonon frequencies, respectively. There are thus at least two approaches to enhance $\lambda_{ep}$: one is to increase $N(E_F)$, and the other is to lower $\langle\omega^2\rangle$ (or, in other words, to soften the lattice). In $Mo_5Si_{3-x}P_x$, $N(E_F)$ is enhanced by electron doping, which shifts $E_F$ to a peak in DOS. The lattice has been effectively softened too, as evidenced by the large decrease of $\Theta_D$. These two factors together give rise to the strong electron-phonon coupling, and should be responsible for the emergence of superconductivity in $Mo_5Si_{3-x}P_x$. The reason why P doping softens the lattice so much (unlike Re doping in $Mo_{5-x}Re_xSi_3$, which did not change $\Theta_D$ much [24]) is definitely worth further studies. Theoretical calculations of the phonon dispersions of $Mo_5Si_{3-x}P_x$ will be illuminating. In particular, large phonon linewidths or soft modes can be expected.

As for the topological properties, although our results suggest that P doping may not be beneficial to the TSSs, there are other ways to chase for topological superconductivity in this system. We notice that the band gap between bands 156 and 158 is topologically nontrivial with strong topological indexes of (1;000) (see Figure 5(f)). Therefore, TSSs are likely to survive and topological superconductivity may be realized, if superconductivity can be induced by hole doping in $Mo_5Si_3$. Another strategy is to fabricate $Mo_5Si_3/Mo_5Si_{1.5}P_{1.5}$ heterojunctions to see whether TSS emerges from the proximity effect. Given the fact that high-quality $Mo_5Si_3$ single crystals are readily available [29], $Mo_5Si_{3-x}P_x$ serves as a suitable platform to observe the possibly existing Majorana zero modes, which will be an intriguing topic in future ARPES or STM studies.

Lastly, we should mention that the $T_c$ of 10.8 K and $\mu_0H_{c2}(0)$ of 14.56 T are fairly high for a pseudobinary compound. For example, both of them are slightly higher than those in the commercial superconductor NbTi ($T_c$ = 9.6 K, $\mu_0H_{c2}(0)$ = 14 T) [40]. Most absorbingly, $Mo_5Si_{3-x}P_x$ shares much in common with $A$15 superconductors such as $V_3Si$ or $Nb_3Sn$. For instance, they have similar $T_c$s, close coupling strength, and all of them show upwards concave features in normal state $\rho(T)$ [30, 41]. Recent theoretical studies suggested nontrivial band topologies in some of the $A$15 superconductors too [42]. Compared to the well-known $A$15 superconductor family, the $W_5Si_3$-type superconductor family, which $Mo_5Si_{3-x}P_x$ belongs to, has not been studied in depth. Currently, the family contains about ten members. Most of them superconduct below 4 K, with the maximal $T_c$ of 5.8 K observed in $Mo_3Re_2Si_3$ [24, 43-48]. Our study almost doubles this maximum, making it comparable with those in the $A$15 family. $W_5Si_3$-type compounds are hence a fertile ground to be explored for new superconductors.

**CONCLUSIONS**

To summarize, we discover that P doping introduces superconductivity in non-superconducting $Mo_5Si_3$, which hosts a nontrivial band topology. $T_c$ increases with the doping level, reaching 10.8 K in $Mo_5Si_{1.5}P_{1.5}$. $Mo_5Si_{1.5}P_{1.5}$ is a type-II, fully gapped superconductor with strong electron-phonon coupling, as evidenced by the large values of $\Delta C_e/\gamma T_c$, $\Delta_0/k_BT_c$, and $\lambda_{ep}$. $\mu_0H_{c1}(0)$ and $\mu_0H_{c2}(0)$ for $Mo_5Si_{1.5}P_{1.5}$ are 105 mT and 14.56 T, respectively. According to first-principles calculations, the large electron-phonon coupling is related to the increase of $N(E_F)$. $Mo_5Si_{1.5}P_{1.5}$ sets a new record of $T_c$ in $W_5Si_3$-type superconductors. Compared to the previously reported $Mo_3Re_2Si_3$ superconductor [24], or the widely used commercial superconducting material NbTi [40], the higher $T_c$ and inexpensive raw materials of $Mo_5Si_{3-x}P_x$ make future applications promising. We point out that the superconducting properties of $Mo_5Si_{3-x}P_x$ are very similar with those in the $A$15 superconductors. Our findings suggest that the $T_c$ levels in $W_5Si_3$-type superconductors can be comparable with the $A$15 superconductors. Novel superconductors with higher $T_c$ values can be anticipated in $W_5Si_3$-type structural family. In particular, superconductivity, or even topological superconductivity, could be achieved through carrier doping, whose effectiveness has been testified in our study.

**Acknowledgements** The authors acknowledge financial support by the National Key Research and Development of China (Grant Nos. 2018YFA0704200, 2021YFA1401800, 2018YFA0305602, and 2017YFA0302904), the

National Natural Science Foundation of China (Nos. 12074414, 12074002, and 11774402), and the Strategic Priority Research Program of Chinese Academy of Sciences (Grant No. XDB25000000).

**Author contributions**     Sun JN conceived the project; Ruan BB and Sun JN synthesized the samples and did most of the measurements; Chen Y, Gu YD and Yang QS assisted in some of the measurements; Ruan BB carried out the theoretical calculations and wrote the paper with supports from Zhou MH, Ma MW and Zhao K; Chen GF, Shan L and Ren ZA reviewed the original manuscript; Ren ZA supervised the project. All authors contributed to the general discussion.

**Conflict of interest**     The authors declare that they have no conflict of interest.

**Supplementary information**     The relaxed lattice parameters from DFT compared with the experimental ones, comparison of VCA and supercell results, SEM image and elemental mapping of $Mo_5Si_{1.5}P_{1.5}$, and the subtraction of $Mo_3P$ contribution from the raw data.

**Table 1** Crystallographic parameters, measured P contents (EDX), and superconducting $T_c$ of $Mo_5Si_{3-x}P_x$ ($0 \leq x \leq 2.0$).

| parameter | $x = 0$ | $x = 0.5$ | $x = 1.0$ | $x = 1.2$ | $x = 1.3$ | $x = 1.5$ | $x = 1.6$ | $x = 2.0$ |
|---|---|---|---|---|---|---|---|---|
| measured P content $x$ | 0 | 0.46(5) | 0.96(9) | 1.14(12) | 1.20(13) | 1.43(14) | 1.49(16) | 1.75(18) |
| $a$ (Å) | 9.6444(1) | 9.6128(1) | 9.5871(1) | 9.5627(1) | 9.5555(1) | 9.5407(1) | 9.5332(1) | 9.4956(2) |
| $c$ (Å) | 4.9063(1) | 4.9200(1) | 4.9345(1) | 4.9447(1) | 4.9492(1) | 4.9582(1) | 4.9674(1) | 4.9856(2) |
| $x_{Si2}$ [a] | 0.1680(2) | 0.1657(3) | 0.1656(3) | 0.1663(3) | 0.1662(2) | 0.1666(1) | 0.1655(4) | 0.1650(4) |
| $x_{Mo2}$ | 0.07694(6) | 0.07684(8) | 0.0767(1) | 0.0771(1) | 0.0770(1) | 0.07630(4) | 0.0767(1) | 0.0770(1) |
| $y_{Mo2}$ | 0.22356(7) | 0.2232(1) | 0.2226(1) | 0.2221(1) | 0.2219(1) | 0.22114(5) | 0.2218(2) | 0.2214(2) |
| $R_p$ | 1.38% | 2.10% | 2.11% | 2.15% | 1.96% | 0.94% | 1.94% | 2.07% |
| $R_{wp}$ | 2.32% | 2.89% | 2.86% | 2.92% | 2.62% | 1.34% | 2.52% | 2.75% |
| $\chi^2$ | 2.48 | 2.51 | 2.08 | 2.24 | 1.74 | 1.56 | 1.76 | 1.74 |
| $Mo_3P$ weight fraction [b] | 0 | 3.4% | 9.9% | 9.4% | 6.0% | 12.3% | 20.6% | 30.2% |
| $T_c$ from $\rho(T)$ (K) [c] | — | 6.77 | 8.70 | 10.09 | 10.42 | 10.74 | 10.74 | 10.70 |
| $T_c$ from $\chi(T)$ (K) | — | 7.01 | 9.26 | 10.18 | 10.40 | 10.71 | 10.54 | 10.71 |

a) Wyckoff positions: Si1 (4$a$), Si2 (8$h$), Mo1(4$b$), Mo2(16$k$). P randomly takes the Si1 and Si2 sites. $x_{Si1} = y_{Si1} = z_{Si1} = z_{Si2} = x_{Mo1} = z_{Mo2} = 0$. $y_{Si2} = x_{Si2} + 0.5$. $y_{Mo1} = 2z_{Mo1} = 0.5$. b) Determined from XRD refinements. c) Determined from the midpoint of superconducting transition in $\rho(T)$.

**Table 2** Superconducting parameters of $Mo_5Si_{1.5}P_{1.5}$. The thermodynamic parameters of $Mo_5Si_3$ and $Mo_5Si_2P$ are also listed for comparison.

| parameter | unit | $Mo_5Si_{1.5}P_{1.5}$ | | |
|---|---|---|---|---|
| $T_c^{onset}$ | K | 10.80 | | |
| $T_c^{zero}$ | K | 10.70 | | |
| $\mu_0H_{c1}(0)$ | mT | 105 | | |
| $\mu_0H_{c2}(0)$ | T | 14.56 | | |
| $\mu_0H_c(0)$ | T | 0.75 | | |
| $\xi_{GL}$ | nm | 4.75 | | |
| $\lambda_{GL}$ | nm | 70.5 | | |
| $\kappa$ | — | 14.8 | | |
| parameter | unit | $Mo_5Si_3$ | $Mo_5Si_2P$ | $Mo_5Si_{1.5}P_{1.5}$ |
| $\beta$ | mJ mol$^{-1}$ K$^{-4}$ | 0.054 | 0.24 | 0.50 |
| $\gamma$ | mJ mol$^{-1}$ K$^{-2}$ | 19.80 | 25.10 | 37.23 |
| $\Theta_D$ | K | 659 | 404 | 314 |
| $\lambda_{ep}$ | — | — | 0.69 [c] | 1.15 [d] |
| $\Delta C_e/\gamma T_c$ | — | — | 2.03 | 2.06 |
| $\Delta_0/k_BT_c$ | — | — | 2.03 | 2.14 |
| $N(E_F)$ [a] | eV$^{-1}$ f.u.$^{-1}$ | — | 6.30 | 7.30 |
| $N'(E_F)$ [b] | eV$^{-1}$ f.u.$^{-1}$ | 4.13 | 7.24 | 8.04 |

a) Experimental value calculated from $\gamma$. b) Theoretical value from DFT calculations. c) Estimated from Equation (4). d) Estimated from Equations (1) and (2).

**Figure 1** (a) Crystal structure of $Mo_5Si_{3-x}P_x$. (b) XRD patterns of $Mo_5Si_{2.5}P_{0.5}$ and $Mo_5Si_{1.5}P_{1.5}$ with their Rietveld refinements. (c) Evolution of lattice parameters $a$ and $c$ upon P doping. (d) The composition $Mo_5Si_mP_n$ determined from EDX measurements. (e) XRD patterns of $Mo_5Si_{3-x}P_x$ ($0 \leq x \leq 2.0$), with a zoom-in of the (141) peak shown in (f).

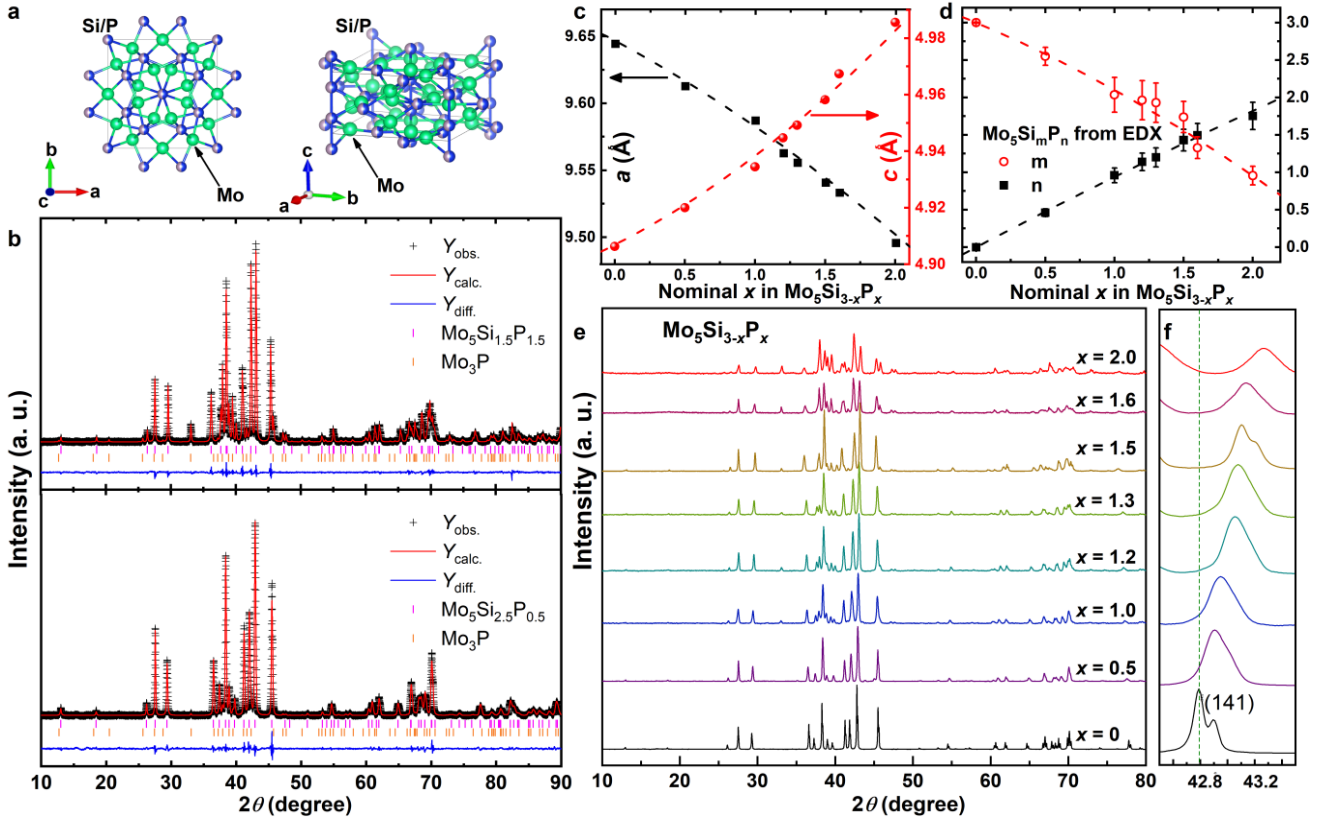

**Figure 2** (a) Temperature dependence of resistivity of $Mo_5Si_{3-x}P_x$ ($0 \leq x \leq 2.0$) under zero magnetic field. (b) Zoom-in of the datasets in (a) below 15 K. (c) DC magnetic susceptibility of $Mo_5Si_{3-x}P_x$ ($0 \leq x \leq 2.0$). (d) Evolution of $T_c$, as well as RRR, with regard to the P doping content $x$.

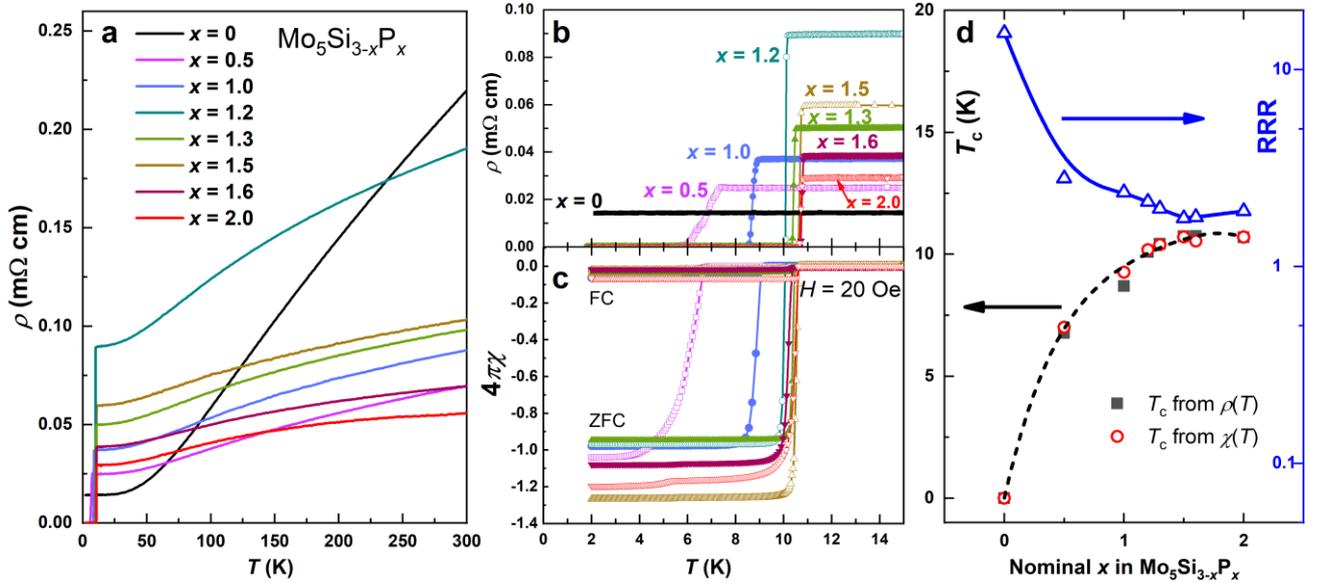

**Figure 3** (a) Superconducting transition on $\rho(T)$ of $Mo_5Si_{1.5}P_{1.5}$ under magnetic fields up to 15.5 T. (b) Isothermal magnetization curves of $Mo_5Si_{1.5}P_{1.5}$ at low temperature. (c) Superconducting hysteresis loop of $Mo_5Si_{1.5}P_{1.5}$ at 2 K. (d) Temperature dependence of the upper and lower critical fields of $Mo_5Si_{1.5}P_{1.5}$.

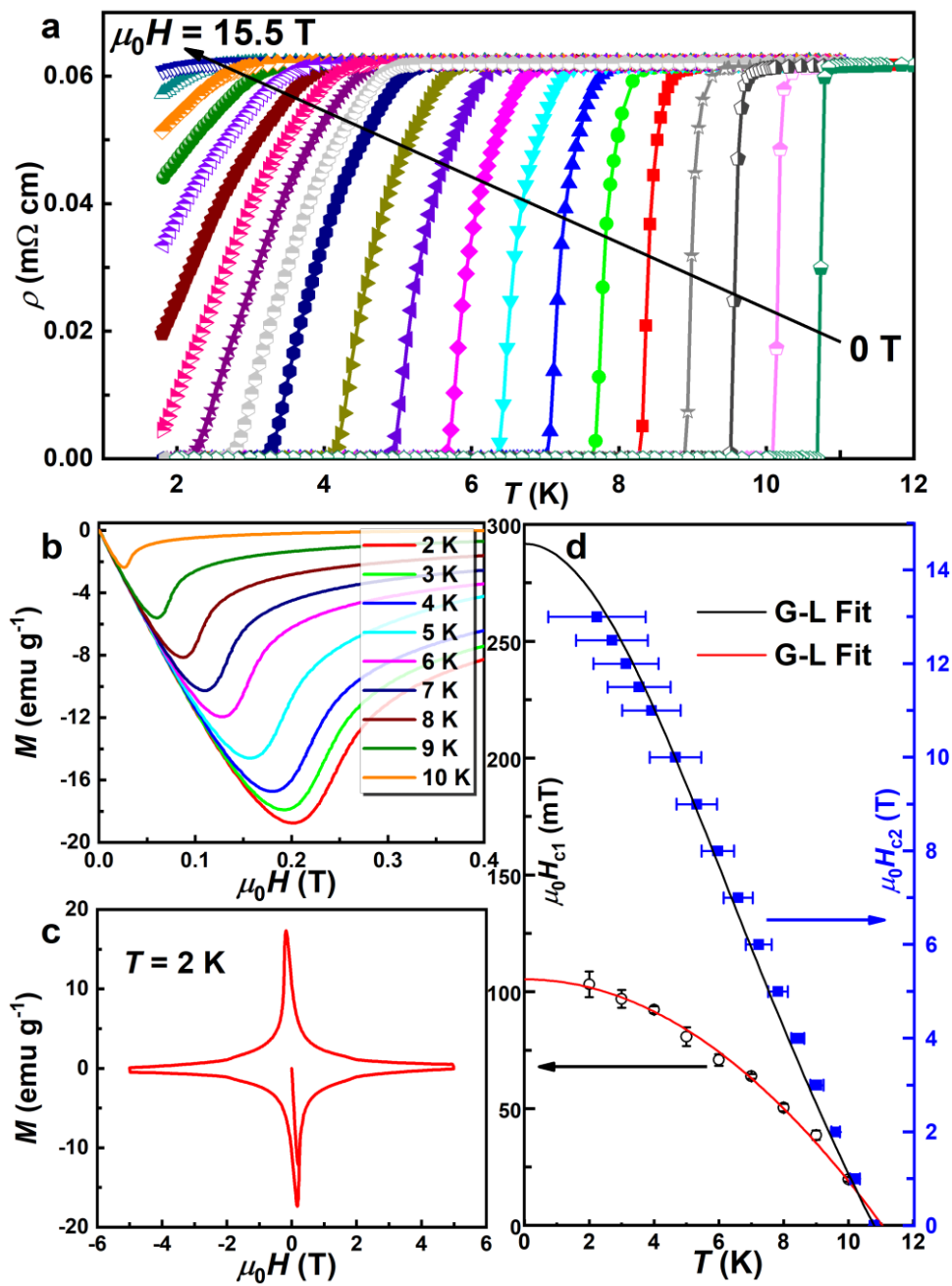

**Figure 4** (a) Temperature dependence of heat capacity of $Mo_5Si_{3-x}P_x$ ($x$ = 0, 1.0, 1.5) under zero magnetic field. The dash lines are fits to the normal state data with the Debye model. (b) Electronic contribution of heat capacity of $Mo_5Si_2P$ and $Mo_5Si_{1.5}P_{1.5}$. The data below $T_c$ are fitted with the $\alpha$ model, shown as the dash lines. Notice the data for $x$ = 1.0 and 1.5 in (a) and (b) have been corrected by subtracting the $Mo_3P$ contributions. The kinks at around 5.6 K are the residual signals from $Mo_3P$ impurity.

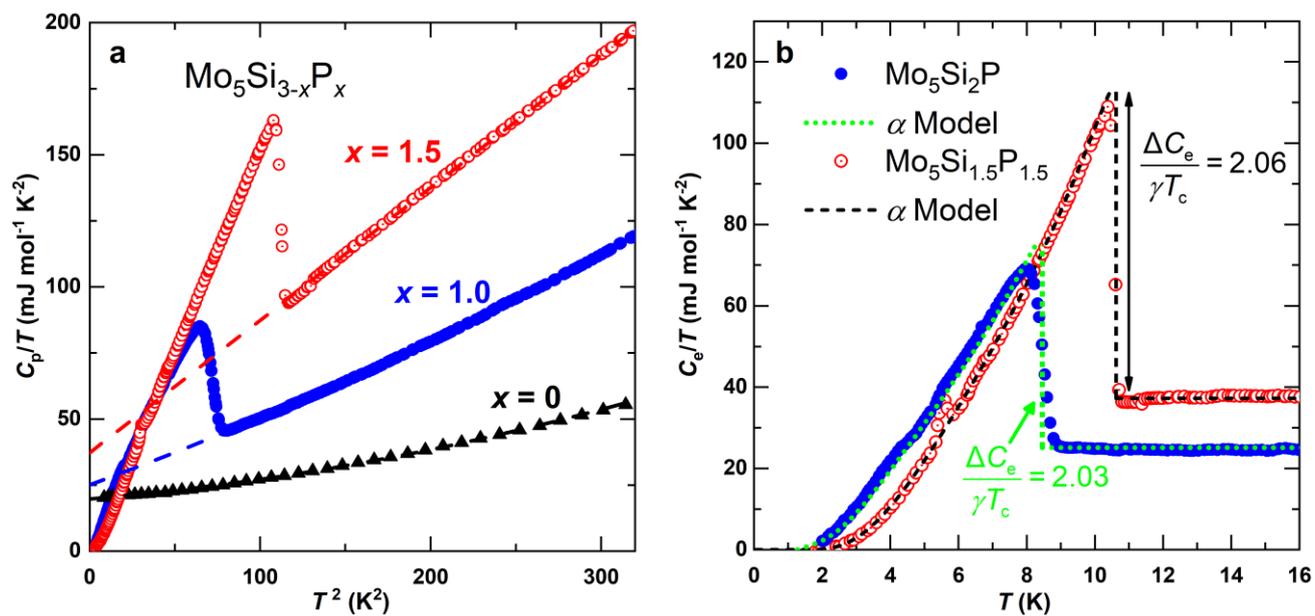

**Figure 5** (a) Band structure of $Mo_5Si_3$ without SOC. (b) Band structure of $Mo_5Si_3$ with SOC. (c) Band structure of $Mo_5Si_2P$ without SOC. (d) Band structure of $Mo_5Si_{1.5}P_{1.5}$ without SOC. The shadowy boxes in (a)–(d) emphasize the flat band dispersions. (e) Brillouin zone of $Mo_5Si_{3-x}P_x$, with high symmetry points labeled. (f) The $Z_2$ topological invariants of $Mo_5Si_3$ (with SOC) for the bands near $E_F$. The band numbers correspond to those in (b). (g) Evolution of DOS of $Mo_5Si_{3-x}P_x$ ($x$ = 0, 1.0, 1.5) upon P doping.

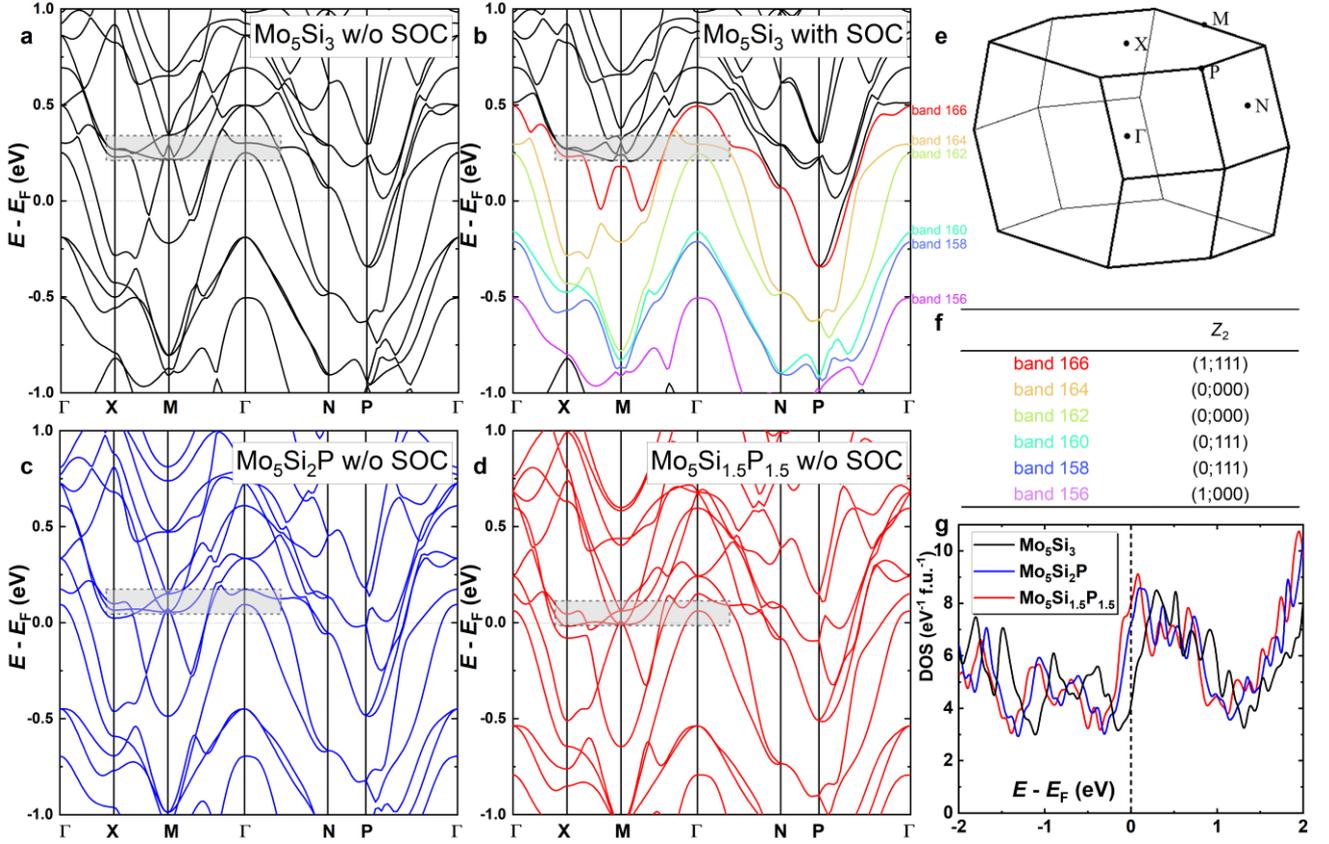